\documentclass[a4paper]{article}
\usepackage[utf8]{inputenc}
\usepackage[T1]{fontenc}
\usepackage{amsmath}
\usepackage{amsfonts}
\usepackage{amssymb}
\usepackage{algorithmicx}
\usepackage{algpseudocode}
\usepackage{algorithm}
\usepackage{natbib}
\usepackage{color}

\newcommand{\De}[3][]{\frac{\partial^{#1} #2}{\partial {#3}^{#1}}}

\newcount\eLiNe\eLiNe=\inputlineno\advance\eLiNe by -1
\newcommand{\xor}{\oplus}
\renewcommand{\eqref}[1]{(\ref{Eq:#1})}

\usepackage{ifthen}
\usepackage{xspace}

\newcommand{\eq}[2][]{%
	\ifthenelse{\equal{#1}{}}{%
		\begin{equation}
		#2%
		\end{equation}%
	}{%
		\begin{equation}\label{Eq:#1}%
		#2%
		\end{equation}%
	}%
}
\newcommand{\meq}[2][]{%
	\ifthenelse{\equal{#1}{}}{%
		\begin{equation}%
		\begin{split}%
		#2%
		\end{split}%
		\end{equation}%
	}{%
		\begin{equation}\label{Eq:#1}%
		\begin{split}%
		#2%
		\end{split}%
		\end{equation}%
	}%
}

\title{Toward a Boundary Regional Control Problem for Boolean Cellular Automata}

\author{Franco Bagnoli$^{(1)}$,
        Samira El Yacoubi$^{(2)}$,
        Ra\'ul Rechtman$^{(3)}$\\[1cm]
\begin{minipage}{0.9\columnwidth}
                 \raggedright
\begin{enumerate}
    \item 
              Dept. Physics and Astronomy \& CSDC, Universit\`a di Firenze\\
              via G. Sansone 1, 50019 Sesto Fiorentino (FI), Italy\\
              Also INFN, sez. Firenze and ISC-CNR, Firenze.
              \texttt{franco.bagnoli@unifi.it}           \\    
\item          
              Team Project IMAGES\_ESPACE-Dev,  \\
              UMR 228 Espace-Dev IRD UM UG UR\\
              University of Perpignan Via Domitia \\
              52, Avenue de Villeneuve. \\ 
              66860-Perpignan cedex. France.  \\
              \texttt{yacoubi@univ-perp.fr }\\
\item            
              Instituto de Energ\'\i{}as Renovables, Universidad Nacional Aut\'onoma de M\'exico, \\
              Apartado Postal 34, 62580 Temixco, Morelos, Mexico. \\
              \texttt{rrs@ier.unam.mx}
            \end{enumerate}
\end{minipage}
}
\date{2017}

\begin{document}

\maketitle

\begin{abstract}
An important question to be addressed regarding  system control on a time interval $[0, T]$ is whether some particular target state in the configuration space   is reachable from a given initial state. When the target of interest refers only to a portion of the spatial domain, we speak about regional analysis. Cellular Automata (CA) approach have been recently promoted for the study of  control problems on spatially extended systems for which the classical approaches cannot be used. An interesting problem  concerns the  situation where the subregion of interest is not  interior to the domain but a portion of its boundary. In this paper we address the problem of regional controllability of cellular automata via boundary actions, i.e., we investigate the characteristics of a cellular automaton so that it can be controlled inside a given region only acting on the value of sites at its boundaries. 
 
\end{abstract}

\catcode`\@=\active
\def@#1{\boldsymbol{#1}}

\section{Introduction}

Cellular Automata (CA) are spatially extended systems that are widely used for modelling various problems ranging from physics to biology, engineering, medicine,  ecology and economics~\citep{ACRI2002,ACRI2004,ACRI2006,ACRI2008,ACRI2010,ACRI2012,ACRI2014,ACRI2016}. An ultimate understanding of such systems gives one the ability to control them in order to achieve desired behaviour.

CA are particularly suitable for simulating biological systems that are normally highly non-linear and better described in terms of discrete units rather than by means of partial differential equations (PDE's)  \citep{Kauffman,Diversity,CAModelingBiologicalPatterns,CAApproachBioModeling}.  In these cases one is interested in finding the CA that best models the problem and in studying the emerging patterns. 

The advantage of using a model with respect to real experiments is that of being able to explore a wide range of parameters and that of  performing measurements that are impossible in real systems. However, one is always concerned with the discrepancies between the model and the real system, differences that can amplify and lead to a complete disagreement between the experimental reality and the model. 

An alternative approach with such a modelling problem was addressed in ~\citet{controlPRE}. If one has at his/her disposal experimental data and assumes that the numerical model is a good but not perfect approximation, one is confronted with the problem of synchronizing the model with the data. From a theoretical point of view, this control problem is equivalent to a ``master-slave'' control problem~\citep{masterslave}. Where is  it more convenient to perform measurements on the master system and apply them to the slave one in order to get their synchronization with the minimum effort (\textit{i.e.}, the optimal control)? 

The results depend on the CA update rule. Clearly, CA that go into a unique final state are trivially easy to control, but they are also extremely unlikely to be a good model for any problem. CAs showing multiple attractors or chaotic behaviour are more interesting. Among them, the unexpected result of our previous study~\citep{controlPRE} is that chaotic ones are easier to control, since it is sufficient to block the spreading of the ``difference'' among the master and the slave and wait for an ``automatic'' synchronization.  This is due to two factors: the ``exploration'' of the configuration space by the chaoticity of the CA and the discreteness of the state variables that makes the synchronized state stationary even if it is unstable with respect to finite perturbations~\citep{LyapunovCA}. 

The problem that we want to address here is that of forcing the appearance of a given pattern inside a region by imposing a suitable set of values to the sites that surround that region. In general, control problems have to be addressed splitting the problem into the \emph{measurability} issue and the actual \textit{controllability} one. The first topic relates to the problem of actually being able to measure some quantity in the  system under investigation. We skip this problem assuming  being able to measure the instantaneous state of the system at will, and we focus on the second problem: that of driving a system into a given state acting only on the periphery of the given region.

This problem is related to the so-called  regional  controllability introduced by \citet{Zerrik}, as a special case of output controllability~\citep{Lions,Russell}. The regional control problem consists in achieving an objective only on a subregion of the domain when some specific  actions are exerted on the system,  in its domain  interior or on its boundaries.   This  concept has been studied by means of partial differential equations. Some  results on the action properties (number, location, space distribution) based on the rank condition  have been obtained depending on the target region and its geometry,  see for example \citet{Zerrik} and the references therein. 

Regional controllability has also been studied using CA models. In \citet{Samira},  a numerical approach based on genetic algorithms has been developed  for a class of additive CA in both 1D and 2D cases.   In \citet{regionalca},  an interesting theoretical study has been carried out for 1-D additive real valued CA where the effect of control is given through an evolving neighbourhood and a very sophisticated state transition function. 
However the study did not provide a real insight in the regional controllability problem and the  theoretical results could not be exploited for other works. In the present article, we  aim at  introducing a general framework for regional control problem by means of CA using the concept of Boolean derivatives.  It focuses on boundary  control and takes into consideration only deterministic one-dimensional CA.

We address here two problems. How to force a given configuration to appear in the controlled region regardless of the initial state of the system and how to force a system to follow a given trajectory, again  in the controlled region. All this, only acting on the boundary of the region. 

These are not exactly the same problems of  boundary regional controllability as those considered for PDE's  where the target region of interest is a portion of the domain boundary, so we can label them as boundary reachability (Section~\ref{sec:regcont}) and boundary drivability (Section~\ref{sec:traj}). 

An application could be that of forcing a certain pattern on a biological system. Many problems in modelling biological systems make use of probabilistic CA, and in this case one has to  deeply modify the  approaches described here. This problem  will be the subject of future investigations.  There is however some literature that uses deterministic models derived from lattice gas cellular automata or their mean-field approximation, lattice Boltzmann equations~\citep{CAModelingBiologicalPatterns}. These are deterministic models that can be investigated using the methods described in the following. These models are only defined  in two or three dimensions, for which the methods here illustrated, although still correct in principle,  need some major modification, especially for what concerns the possibility of controlling the whole periphery of the region of interest. So for the moment this article should be considered as a first step toward the study of a rather complex problem. 

The sketch of the paper is the following. In Section~\ref{sec:defs} we present some definitions and the concept of Boolean derivatives, which are the analogous of standard derivatives for discrete systems. In Section~\ref{sec:regcont} we address the actual problem of boundary reachability~\citep{regional} and present some theorems and conjectures about the classification of controllable CA. In order to address the control problem numerically, in Section~\ref{sec:preim} we present a method for generating all pre-images of a target configuration for a given CA. In Section~\ref{sec:traj} we address the problem of boundary drivability.  Finally, conclusions are drawn in the last section. 

\section{Definitions and Boolean derivatives}\label{sec:defs}

A CA is defined by a set of $N$ individual automata, sitting on the nodes of a graph, that defines the connections among a node and its neighbours. Each individual automaton can assume one out of a set of states and owns a transition rule (generally the same for all automata) for the updating of the states according with the values of those of neighbours. The graph is defined by an adjacency matrix $a$ such that $a_{ij}=1$ if $j$ is a neighbour of $i$ and zero otherwise. The adjacency matrix might not to be symmetric. 

A lattice is a graph invariant by translation, \textit{i.e.}, $a$ commutes with the shift matrices, which in one dimension are $S^{(L)}_{ij}=[[j=i+ 1]]$ and $S^{(R)}_{ij}=[[j=i-1]]$, where $[[\cdot]]$ is the truth function, a generalization of the Kronecker delta which takes values 1 if $\cdot$ is true and zero otherwise, and $L$ ($R$) stands for the left (right) shift. Operations on the indices are assumed to be periodic modulo $N$. 

The (input) connectivity degree of a node $i$ is defined by $k_i = \sum_j a_{ij}$. In the following we shall consider one-dimensional lattices with homogeneous degree $K=2k+1$. We shall call $k$ the \textit{range} of the automaton. 

Let us denote the state of automaton $i$ at time $t$ by $s_i^t$. In the simplest version, which is the one considered here, the state can only take  two values, one and zero (Boolean automata). 
The state $@n_i^t$ of the neighbourhood of a site $i$ at time $t$ is given by the set of the states of the  $K$  connected sites $@n_i^t = \{s_{j}^t | a_{ij}=1\}$. For the one-dimensional lattice,  $@n_i^t = \{s_{i-k}^t, \dots, s_i^t,\dots, s_{i+k}^t\}$.

The evolution of the CA is given by the parallel application of the updating rule $f$,
\eq[evolution]{
	s_i^{t+1} = f(@n_i^t).
}
The function $f$ is a Boolean function of $K$ Boolean arguments.
Since all variables take Boolean values, it is possible to read the set of state of the neighbours 
\[
\{s_{i-k}^t, \dots, s_i^t,\dots, s_{i+k}^t\}\equiv\{x_0, x_1, \dots x_{K-1}\}
\]
 as a base-two number $X=\sum_{j=0}^{K-1} x_j2^j$;  $0\le X< 2^K$.

In order to indicate a CA in a compact way, one can use the Wolfram's notation~\citep{Wolfram}. Just consider the set of values that the function takes for all possible configurations of the neighbourhood, ordered as a number in base two
\eq[wolframcode]{
	\{  f(2^K-1), \dots, f(1), f(0)\}
} 
and read it as a base-two number $\mathcal{F}=\sum_{j=0}^{2^K-1} f(j)2^j$;  $0 \le \mathcal{F}<2^{2^K}$. This notation is actually useful only for \emph{elementary} (Boolean, $K=3$) automata. 

It is possible to define an equivalent of the usual derivatives for such discrete systems~\citep{Vichniac,derivative}. Given a Boolean function 
\eq[booleanfunction]{
	x' = f(x_1, x_2, \dots, x_i, \dots),
}
the Boolean derivative of $x'$ with respect to $x_i$ measures if $x'$ changes when changing $x_i$ is defined as 
\eq{
	\De{x'}{x_i} = f(x_1, x_2, \dots, x_i, \dots) \xor f(x_1, x_2, \dots, x_i\xor 1 , \dots),
}
where $\xor$ is the sum modulus two (XOR operation). The Boolean derivative is one if, given the arguments $\{x_1, x_2, \dots, x_{i-1},x_{i+1} , \dots\}$, $x'$ changes its value whenever $x_i$ does, and is zero otherwise. 

This definition fulfils many of the standard properties of the derivative, and this is particularly evident if one expresses the function $f$ using only AND (multiplication) and XOR operations. For instance, the Boolean derivative of $f(x_1,x_2)=x_1\oplus x_2$ with respect to $x_1$ is
\eq{
	\De{(x_1\xor x_2)}{x_1} = x_1\xor x_2 \xor (x_1\xor 1) \xor x_2 = 1
}
since $x\xor x = 0$. Analogously,  the Boolean derivative of $f(x_1,x_2)=x_1x_2$ with respect to $x_1$ is
\eq{
	\De{(x_1 x_2)}{x_1} = x_1 x_2 \xor (x_1\xor 1) x_2 = x_1x_2 \xor x_1x_2\xor x_2 = x_2. 
}

The derivative can be used to define an analogous of the Taylor series, which, for Boolean functions, is always finite. So for instance, expanding with respect to $x=0$,
\eq{
	f(x) = f(0) \xor f'(0) x = f(0) \xor (f(0)\xor f(1)) x,
}
where $f'$ is the derivative of $f$. It is immediate to verify the above identity by substituting the two possible values of $x$. In general 
\eq{
	f(x\xor y) = f(x) \xor f'(x) y.
}

The derivative also obeys to the chain rule
\meq[chain]{
	\De{f(g(x))}{x}& = f(g(x)) \xor f(g(x\xor 1)) \\
	&= f(g(x)) \xor f(g(x) \xor g'(x)) \\
	&= f(g(x)) \xor f(g(x)) \xor f'(g(x)) g'(x)=  f'(g(x)) g'(x).
}

Since in general a Boolean function depends on many variables $x_0, x_1, \dots$, it is more compact to indicate a derivative as $f^{(i)}$ where $i$,  in base two, indicates which variables are varied for taking the derivative. For instance, 
\meq{
	&f^{(1)}(\dots,x_1, x_0) = \De{f(\dots,x_1, x_0)}{x_0}; \quad  f^{(2)}(\dots,x_1, x_0) = \De{f(\dots,x_1, x_0)}{x_1}; \\
	&f^{(3)}(\dots,x_1, x_0) = \frac{\partial^2 f(\dots,x_1, x_0)}{\partial x_0\partial x_1}; \quad f^{(4)}(\dots,x_1, x_0) = \De{f(\dots,x_1, x_0)}{x_2};  \\
	&\dots
}

Using the Taylor (or McLaurin) expansion, it is evident that every function can be written as a sum (XOR) of polynomials (AND) of the variables, for instance
\eq{
	f(x,y,z) = f_0 \xor f_1 x \xor f_2 y \xor f_3 xy \xor f_4 z \xor f_5 xz \xor f_6 yz \xor f_7 x y z,
}
where $f_i = f^{(i)}(0)$ is the $i$-th derivative of $f$ in zero (the McLaurin coefficient), and $f_0=f(0,\dots,0,0)$. The subscripts $i$ of $f_i$,  in base two, indicate the variables composing the polynomial. 

The function is uniquely identified by the set of values of the derivatives in zero, $\{f_0, f_1, \dots\}$, which therefore constitute an 
alternative to the set of values that the function 
takes for all configurations, $\{f(0), f(1), \dots\}$, used in the Wolfram's notation. 
For instance, the elementary CA W150 $=\{1,0,0,1,0,1,1,0\}$, \textit{i.e.}, $x_0\xor x_1\xor x_2$, can be identified by the McLaurin coefficients $\{0,1,1,1,0,0,0,0\}$.

The XOR function can be expressed in terms of other Boolean functions, as for instance $x\xor y = (x \wedge \overline{y}) \vee (\overline{x} \wedge y)$, where $\wedge$ is the AND and $\vee$ the OR operation, and the line marks the negation (NOT) operation. In this way one can recover the standard disjunctive and conjunctive forms of Boolean functions, although using the Taylor expansion with Boolean derivatives,  one can express any function  using only two operations, XOR and AND~\citep{derivative}. 

\section{Boundary reachability for cellular automata} \label{sec:regcont}
Let us denote by $c=\{c_1$, $\dots$, $c_{W-1}\}$ the state of the region to be controlled, see Fig.~\ref{fig:regional}. The idea is to impose the state of 
all or some of the boundary sites, denoted as $\ell_{-1}^t$ (left sites) and $r_{W}^t$ (right sites), so that the states of cells in the region   at time $T$, $c^T=\{c_0^T$, $\dots$, $c_{W-1}^T\}$ be equal to the desired ones  $q=\{q_0$, $\dots$, $q_{W-1}\}$, for every initial condition $c^0=\{c_0^0$, $\dots$, $c_{W-1}^0\}$. If the control can be done in the minimum time $T=W/(2k)$ (the minimum time for  letting a signal coming from the $\ell$ or $r$ regions to reach all sites), the control is said to be optimal. 

The problem that we address is the following: given a one-dimensional lattice and a Boolean function $f$ with a neighbourhood of size $K=2k+1$, what are the conditions on the function so that a region of arbitrary size $W$ can be driven to a given state $q$ in a time $T$ regardless of its initial state $c$, only acting on its boundary? Notice that the boundary  has to be of width equal to the range  $k$, so that all sites inside the region to be controlled depend either on the previous state of sites inside the same region, or on the state of sites in the boundaries.  

An interesting observation allows  to switch from the boundary control to an initial-value control. Let us consider an initial-value problem as  shown in Fig.~\ref{fig:initial}, where we only fix an appropriate number of cells  $\ell^0=\{\ell_i^0\}$ at the left and $r^0=\{r_j^0\}$ at the right of the region to be controlled, and then let the system evolve. If, in this way, we are able to obtain the desired state 
$q$ in the central region at time $t$, it is sufficient to apply the sequence of $\ell_{-1}^t$ and  $r_{W}^t$ to obtain the desired boundary control. So we can limit our study to this initial value problem. 

\begin{figure}[t]
	\begin{center}
		\begin{tabular}{cc|cccc|cc}
			$\dots$& $\ell_{-1}^0$ & $c_0^0$ & $c_1^0$ & $\dots$ & $c_{W-1}^0$ & $r_W^0$ &  $\dots$\\[.5cm]
			$\dots$& $\ell_{-1}^1$ & $c_0^1$ & $c_1^1$ & $\dots$ & $c_{W-1}^1$ & $r_W^1$ &  $\dots$\\[.5cm]
			& $\vdots$& $\vdots$& $\vdots$& & $\vdots$& $\vdots$\\[.5cm]
			$\dots$& $\ell_{-1}^T$ & $q_0$ & $q_1$ & $\dots$ & $q_{W-1}$ & $r_W^T$ &  $\dots$\\
		\end{tabular}
	\end{center}
	\caption{\label{fig:regional} Regional control for $k=1$.}
\end{figure}

We can restate the control problem in the following way. Each site in the target configuration $c_i^T$ depends on a set of states at time $0$
\eq{
	c_i^T= F(\ell^0,c^0,r^0),
}
where we have indicated schematically the values of the set of sites of the region to be controlled ($c$) and those at its left ($\ell$) and  right ($r$) . The function $F$ is given by the repeated application of the local CA updating function $f$ for $T$ time steps. By using the chain rule, Eq.~\eqref{chain}, one can obtain the dependence of $c_i^T$ from the sites at time 0
\eq{
	\De{c_i^T}{\ell_j^0},\qquad  \De{c_i^T}{c_n^0},\qquad \De{c_i^T}{r_m^0}. 
}

We want to be able to impose an arbitrary configuration to $c_i^T$ for any given $c_j^0$ by changing the values of $\ell^0$ or $e^0$. This means that, for every initial configuration $c^0$, there should exist a set of $\ell^0$ and $r^0$ such that there exists at least one $\ell_n^0$ or $r_m^0$ such that 
\eq{
	\De{c_i^T}{\ell_n^0} =1 \qquad \text{or} \qquad \De{c_i^T}{r_m^0}=1. 
}

\begin{figure}[t!]
	\begin{center}
		\begin{tabular}{cccc|cccc|cccc}
			$\dots$&$\ell_{-3}^0$ & $\ell_{-2}^0$ & $\ell_{-1}^0$ & $c_0^0$ & $c_1^0$ & $\dots$ & $c_{W-1}^0$ & $r_W^0$ & $r_{W+1}^0$& $r_{W+2}^0$& $\dots$\\[.5cm]
			$\dots$&$\ell_{-3}^1$ & $\ell_{-2}^1$ & $\ell_{-1}^1$ & $c_0^1$ & $c_1^1$ & $\dots$ & $c_{W-1}^1$ & $r_W^1$ & $r_{W+1}^1$& $r_{W+2}^1$& $\dots$\\[.5cm]
			& $\vdots$& $\vdots$& $\vdots$& $\vdots$& $\vdots$&& $\vdots$ & $\vdots$ & $\vdots$ & $\vdots$\\[.5cm]
			$\dots$&$\ell_{-3}^T$ & $\ell_{-2}^T$ & $\ell_{-1}^T$ & $q_0$ & $q_1$ & $\dots$ & $q_{W-1}$ & $r_W^T$ & $r_{W+1}^T$& $r_{W+2}^T$& $\dots$\\
		\end{tabular}
	\end{center}
	\caption{\label{fig:initial} Initial-value control}
\end{figure}

As expected, any linear CA (sum of degree-one polynomials, plus eventually a constant) is controllable~\citep{Samira}, unless it only depends on the previous value of the same site.

One of the main results  is that  a CA  is controllable for reachability if its updating function  is linear with respect to at least one of the peripheral sites (peripherally linear).  A function 
\eq{
	x_i' = f(x_{i-k}, x_{i-k+1},\dots, x_{i}, \dots,x_{i+k-1}, x_{i+k}),
}
is peripherally linear if  
\eq{
	\De{x'}{x_{i+k}} =1 \qquad \text{or} \qquad \De{x'}{x_{i-k}} =1.
}
This means that 
\meq{
	x_i' & = g_\ell(x_{i-k}, x_{i-k+1},\dots, x_{i}, \dots, x_{i+k-1}) \xor  x_{i+k} \\
	& \qquad \text{or} \qquad x_i' = x_{i-k} \xor g_r(x_{i-k+1},\dots, x_{i}, \dots,x_{i+k-1}, x_{i+k}) ,
}
for suitable functions $g_\ell$  and $g_r$, .

Let us consider the case of a function which is peripherally linear on the right.  In this case, the value of $x_{i+k}$ can force the value of $x'$ for every set of $\{x_{i-k}, x_{i-k+1},\dots, x_{i}, $ $\dots, x_{i+k-1}\}$. This means that $c_0$ at time ${W/(k+1)}$,  ($c_0^{W/(k+1)}$) depends linearly on $r_{W}^0$. Setting this value so that $c_0^{W/(k+1)}=q_0$, we can proceed to fix the value of 
$c_1^{W/(k+1)} = q_1$ by using its linear dependence on  $r_{W+1}^0$ and so on. So, peripherally linear CAs are controllable. For these CAs, one only needs to apply the control to one side, therefore in this case it is possible to control a region from just one boundary. However, they are not optimal.

The double peripherally linear  CAs are optimally controllable for reachability. In this case, one divides the set of target sites in two (left and right half), and uses leftmost or rightmost sites to fix the value of the target region at the minimum time. Clearly, fully linear CAs like W150 or W90 are optimally  controllable.

A CA is not controllable if it depends in a multiplicative way on the previous value of the same site so that it can ``pin'' the value of that polynomial to zero. For instance, if 
\meq{
	x_i'& = f(x_{i-k}, x_{i-k+1},\dots, x_{i}, \dots,x_{i+k-1}, x_{i+k}) \\
	&= x_i g(x_{i-k}, \dots, x_{i-1},x_{i+1} \dots, x_{i+k}),
}
it is impossible to force  $x'_i$ to one,  if $x_i=0$. By composing the transition functions, it is evident that the configuration $\{c_i^0=0\}$ ``pins'' $\{c_i^T=0\}$ and thus this CA is not controllable. 

What about  other CAs?  Since we assume that they are not peripherally linear, the left-most and right-most sites appear in some polynomial together with other sites. By composing the transition functions backward in time, it implies that the sites in the $\ell$ and $r$ regions may appear, in the expression for the sites in the target region, in polynomials in which also sites of the control region appear, for instance 
\eq[cT]{
	c_i^T = \ell_k^0 c_n^0 \xor \ell_m^0 c_p^0 r_u^0,
}
where the symbols on the \textit{r.h.s.} of \eqref{cT} stand for the values of the automata in each region at time $t=0$. 

This means that the configuration in which $c_n^0=0$ and  $c_p^0=0$ cannot be controlled. Notice that the number and size of polynomial terms increases when composing the functions. 

The CA may  depend linearly on some other cell (not at the periphery) and in some cases this is sufficient to make the CA controllable. However, in general this site is also involved in some polynomial of some other site to be controlled, and it may become not available for control. For instance, if 
\eq{
	c_i^T = \ell_k^0 \xor \dots ;\qquad \text{but}\quad c_j^T = \dots \ell_k^0 c_n^0 \dots
}
the state of $\ell_k^0$ may be required to take value one for setting the value of $c_j^T$ to one when $c_n^0=1$, and thus cannot be used to control $c_i^T $. 

So in general, non-peripherally linear CAs are not controllable, with exceptions.  As shown in the following section, some of them obey boundary reachability, and some are not. For the moment we do not have a general criterion for discriminating among them.

\section{Boundary reachability by generating the  preimages of a configuration}\label{sec:preim}

In order to numerically test for boundary reachability, one should in principle, for each initial configuration $C^0=\{c_i^0\}$ and all desired target configurations $Q=\{q_i\}$, search for the existence of a configuration of $\ell^0=\{\ell_i^0\}$ and $r^0=\{r_i^0\}$ that generates the target configuration, this for various sizes of the control region and time $T$. One can save some computational time by reversing the problem, starting from the target configuration and backtracking all pre-images. It is relatively easy to recursively generate all pre-images of a configuration, and thus test that they include all possible antecedents of the control region. If this is not the case, the CA is not controllable in such a time interval. 

The backtracking algorithm we used is discussed in details in \citet{BagnoliAcri2016}, and the complete source in C language can be found in \citet{code}. 
With the use of this program it is easy to compute the boundary reachability of CAs for small regions and limited time span. We consider sizes up to $W=6$ and times up to $T=6$. 

Let us consider the case $K=5$ ($k=2$). We denote the neighbourhood of a given site  as $\{x_{ll}, x_l, x_c, x_r, x_{rr}\}$. As expected, any peripherally linear CA is controllable, like for instance $x_{ll}\xor x_l x_c x_r x_{rr}$, and all double-peripherally linear CAs are optimally controllable, like for instance $x_{ll}\xor x_l x_c x_r \xor x_{rr}$. If the leftmost (or rightmost) site is not involved in any polynomial, the CA is still peripherally linear and thus controllable (although not optimally), like $x_l\xor x_c x_r x_{rr}$.

Let us now consider non-peripherally linear CAs, i.e., CAs whose transition functions is like $x_{ll} x_l \xor x_c \xor x_r x_{rr}$ or $x_{ll} x_{rr} \xor x_c \xor x_r x_{l}$. As reported in Table~\ref{tab:mu}, it is not evident which  structure is responsible for controllability. An indication may come from some estimation of the chaoticy of the CA (which implies a strong dependence on variation of inputs). It is possible to define the equivalent of the maximum Lyapunov exponent for cellular automata~\citep{lyapPLA,LyapunovCA}, which in principle depends on the trajectory and thus on the configuration. A rough idea of the chaoticity of the CA can be given by the average number of ones in the Jacobian for a random configuration, which is readily computable by evaluating the average value $\mu$ of all first-order derivatives over all possible configurations. We also tried to assign higher weight to peripheral derivatives with respect to central ones ($\mu'$). 

\begin{table}[t]
	\caption{\label{tab:mu} The average number of ones in the derivative ($\mu$) and the weighted version ($\mu'$) for some CAs, compared with the minimal control time $T_c$ for some values of $W$ (width of the controlled region). For some CA the control time (if any) for $N=6$ was longer than the available computational time, so $N$ was reduced (these CAs are assumed not to be controllable). The quantity $T_b(10)$ is the average synchronization time  (over 1000 samples) for a width $W=10$ over $N=1000$ sites. If the synchronization time exceeds 10000 the CA is considered not boundary drivable.  \textbf{LP?} stands for linearly peripheral, \textbf{C?} for controllable and \textbf{BD?} for boundary drivable.}
	\begin{center}
		\begin{tabular}{ccccccccc}
			\textbf{CA} & \textbf{LP?} & $\mathbf{\mu}$ &$\mathbf{\mu}'$ & $T_c$ & $W$ &\textbf{ C?} & $T_b (10)$ & \textbf{BD?}\\
			\hline
			$x_{ll} \xor x_l \xor x_c \xor x_r \xor x_{rr}$ & yes & 5 &  6 & 2 & 6 & yes & $> 10000$ &  no\\
			$x_{ll} \xor x_l  x_c  x_r \xor x_{rr}$ & yes& 2.75 & 4.5 & 2 & 6 & yes & 1200 & yes\\
			$x_l \xor x_c \xor x_r \xor x_{rr}$ & yes& 4 &4 & 2 & 6 & yes & 8 & yes\\
			$ x_l \xor x_c \xor x_r $ & yes& 3 &2 & 3 & 6 & yes & $>10000$ & no\\
			$x_{ll} \xor  x_l  x_c  x_r  x_{rr}$& yes & 1.5 & 2.5 & $3$ & 6 & yes & 5 & yes\\
			$x_{ll} x_{rr} \xor x_l \xor x_c \xor x_r$& no & 4 & 4 & 6 & 6 &yes & 300 & yes \\
			$x_{ll} x_{rr} \xor x_c \xor x_r $& no & 3 & 3 &6 & 4 & no & 150 &yes  \\
			$x_{ll} x_{l} \xor x_c \xor x_r x_{rr} $ & no & 3 & 3 &6& 5 & no & $>10000$ & no\\
			$x_{ll} x_{rr} x_l \xor x_c \xor x_r $ & no & 2.75 & 2.25 & $5$ & 4 & no  & 500 & yes\\
			\hline
			
		\end{tabular}
	\end{center}
\end{table}

The comparison between the values of $\mu$ and $\mu'$ and controllability is reported in Table~1  for some CAs. The method can only say if a small region can be controlled in a limited amount of time. So, one can only state if a control can be found within this time limit. As shown in the Table, the chaotic indicator shows some correlation with the minimum control time $T_c$, but this indicator is not exhaustive, since peripherally linear CA can be easily controlled even when they show relatively low chaoticity indicators, while non-peripherally linear CA may be hard or impossible to control even in the presence of moderately high chaoticity.

\section{Forcing a trajectory (boundary drivability)}\label{sec:traj}

A related, but different problem is that of forcing the driven system to follow a certain trajectory, for instance to maintain for all times the configuration imposed. 

Let us state the problem in a more formal way. Let us denote by $c^t=\{c_0^t$, $\dots$, $c_{W-1}^t\}$ a trajectory of the sites in the target region. 
Is it possible to manipulate the state of the boundaries  $\ell_{-1}^t$ (left sites) and $r_{W}^t$ (right sites) in order to have $c^t$ to be equal to an arbitrary desired trajectory $q^t$, possibly after a transient? 

The general result is that this is not possible, even for linear CA that are controllable in the boundary reachability sense. The proof is quite simple, and it is based on  counting  the available degrees of freedom. In order to force the appearance of a given configuration, we can act on the sites of the initial configuration (\textit{i.e.}, at time $t=0$) in the vicinity of the controlled region. If we want to force the appearance of another (or the same) configuration at the subsequent time steps, we should act similarly on the same region at time $t=1$, but, except for those in a strip of size $k$ at the boundary of such region, the state of all other sites is determined by the previous configuration. So, for a sufficiently large width of the target region, we simply do not have enough ``free sites'' to exert the control and force the appearance of an arbitrary trajectory. 

However, the control is in principle possible if we only want to make the controlled region follow a ``natural'' trajectory (\textit{i.e.}, one which would be  followed in the absence of any control, if the starting configuration were the desired one (and this relates to the reachability problem of Section~\ref{sec:regcont}) and the state of the boundary are those needed to maintain the trajectory. In these terms, the problem can be mapped onto a synchronization task:  take a second  system that is following the desired trajectory (we assume that it exists since we are focussing on natural trajectories). How can we make the target (slave) system synchronize with the second (master) one, only acting on the boundaries?

We therefore consider a master-slave synchronization where $x$, the master, evolves in time and acts on the boundaries of the central region $c$ of size $W$ of $y$, the slave, with the goal that in that region, $y_i=x_i$. 
The order parameter (or observable) is $n(W)$, the number of sites in $C$ where 
$x_i\oplus y_i=1, i\in C$. Complete synchronization is achieved when $n=0$.

All controllable CAs can be trivially forced to follow a natural trajectory, since they can be forced to assume the synchronized configuration in the target region and then they will follow their natural trajectory, with the value of the sites on the boundary copied from the master system.

However, one can ask if a CA can be driven onto a natural trajectory by the simple mechanism of imposing the synchronization over the boundaries of the region, \textit{i.e.}, by synchronizing, for all time steps, the $\ell_i$ and $r_i$ sites of Fig.~\ref{fig:regional}. Let us denote this mechanism with the term ``boundary drivable''. 

We performed numerical experiments on the same rules examined for boundary reachability.
As can be seen in the rightmost two columns of Table~\ref{tab:mu}, boundary reachability does not imply boundary drivability, since one can have examples of both behaviours. A theoretical explanation  is still missing.

\section{Conclusions}\label{sec:conclusions}
We presented the problem of regional controlling  Cellular Automata (CA) only acting on the boundary of a given region (boundary  controllability), and applied it to the problem of driving a target region to assume a desired state (boundary reachability). This problem has proved to be quite hard. We showed that it can be remapped into an initial-value problem and furnished an explicit solution for CA  that are peripherally linear.

We developed a method for numerically  analysing  the pre-images of a given CA configuration and check, for small sizes of the target region and small time intervals, the possibility of boundary reachability. Better algorithms are needed for numerically investigating  the problem in higher dimensions or on graphs and lattices with larger connectivity. 

We also introduced the problem of boundary drivability, where an automaton can be driven to follow a natural trajectory just by synchronizing the sites at the boundaries of a given region.  We showed that boundary reachability and boundary drivability are independent properties. 

We investigated also the possible role of chaoticity indicators, but the conclusions are not definitive. 

The obtained results deal with a specific class of  deterministic Boolean CA and will be extended in the future to  more general cases.

\section*{acknowledgements}
F.B. acknowledges partial financial support from CNR, short term mobility 2016 CUP B52F16001810005.


\end{document}